\documentclass[twocolumn,showpacs,preprintnumbers,amsmath,amssymb,floatfix]{revtex4-2}
\usepackage{graphicx}
\usepackage{dcolumn}
\usepackage{bm}
\usepackage{latexsym}
\usepackage[hang,small,bf]{caption}
\usepackage[subrefformat=parens]{subcaption}
\usepackage{braket}
\usepackage{here}
\usepackage{multirow} 
\newcommand{\be}{\begin{equation}}
\newcommand{\ee}{\end{equation}}
\newcommand{\bea}{\begin{eqnarray}}
\newcommand{\eea}{\end{eqnarray}}

\def\Tr{\text{Tr}} 
\def\nn{\nonumber}

\newcommand{\rcnp}{\affiliation{Research Center for Nuclear Physics, Osaka University, Osaka 567-0047, Japan}}

\begin{document}
\title{Monopoles of the Dirac type and color confinement in QCD\\
- Gauge invariant mechanism -
%
}
\author{Tsuneo Suzuki}
\email[e-mail:]{tsuneo@rcnp.osaka-u.ac.jp}
\rcnp 

\date{\today}

\begin{abstract}
As a color confinement mechanism,  a dual Meissner effect due to Abelian monopoles involved in QCD has been discussed so far in various ways. 
But still there is an important problem unsolved. It is  gauge invariance of the schemes or, in other words,  the reason why \textit{non-Abelian} color confinement is explained by means of the \textit{Abelian} dual Meissner effect. 
Here it is shown that an Abelian method proposed by the present author based on the violation of non-Abelian Bianchi identity (VNABI) can prove $SU(3)$ invariance and explain why color-singlets alone can survive in the confinement phase of QCD in the framework of the \textit{Abelian} dual Meissner effect due to Abelian monopole condensation. This is completely different from the 't~Hooft's Abelian projection scheme which introduces an additional partial gauge-fixing or the idea of Bonati et al. which is based also on VNABI. Bonati et al. say that VNABI 
 is related to the 't~Hooft tensor and the relation 
 can prove
   the gauge invariance of the 't~Hooft's Abelian projection schemes. But the last idea is found to be incorrect. Existence of 
the relation between VNABI and the 't~Hooft tensor 
 alone  can not account for gauge invariance of 
 't~Hooft's
  Abelian projection schemes. 
\end{abstract}

\pacs{12.38.AW,14.80.Hv}

\maketitle
\section{Introduction}
Color confinement in quantum chromodynamics (QCD) is still an important unsolved  problem. As a picture of color confinement, 't~Hooft~\cite{tHooft:1975pu} and Mandelstam~\cite{Mandelstam:1974pi} proposed an interesting idea that the QCD vacuum is a kind of a magnetic superconducting state caused by condensation of magnetic monopoles and  an effect dual to the Meissner effect works to confine color charges. If the dual Meissner effect picture is correct, it is absolutely necessary to find color-magnetic monopoles only from gluon dynamics of QCD. 

An interesting idea to introduce such a monopole in QCD is to reduce $SU(3)$  to an Abelian maximal torus group by a partial (but singular) gauge-fixing~\cite{tHooft:1981ht}.  In $SU(3)$ QCD, the maximal torus group is  $U(1)^2$. Then Abelian magnetic monopoles appear as a topological object at the space-time points corresponding to the singularity of the gauge-fixing matrix. Condensation of the monopoles  causes  the dual Meissner effect with respect to the remaining $U(1)^2$. Numerically,  Abelian projection schemes in the maximally Abelian (MA) gauge~\cite{Kronfeld:1987ri,Kronfeld:1987vd,Suzuki:1992rw, Chernodub:1997ay} and in various unitary gauges~\cite{Sekido:2007mp} seem to support the conjecture. However, $SU(3)$ gauge invariance or gauge-choice independence of this scheme is not explained. Moreover, it is not shown why color singlets alone can survive in the physical state under the Abelian projection scheme.   

In 2010 Bonati et al.~\cite{Bonati:2010tz,Bonati:2010B} 
found an interesting relation showing 
 that the violation of non-Abelian Bianchi identity (VNABI) exists behind the 
 't~Hooft's
 Abelian scenario in various gauges. They claim that monopole condensation is proved to be gauge invariant only from 
the relation. 
Moreover, they
make an  astonishing assertion 
 that the Maximally Abelian (MA) gauge
\cite{Kronfeld:1987ri,Kronfeld:1987vd} which is well known to give us interesting \textit{numerical} results in lattice Monte-Carlo simulations has a special preferred position also \textit{theoretically} among infinite possible gauge-fixing methods. The standpoint proposed in Refs\cite{Bonati:2010tz,Bonati:2010B} is called
as Pisa scheme from now on.

In 2014, the author~\cite{Suzuki:2014wya} found  a general relation that, if the non-Abelian Bianchi identity is broken in QCD, Abelian-like monopoles necessarily appear due to a singularity leading to  non-commutativity with respect to successive partial derivatives without introduction of additional gauge-fixing as done in 
't~Hooft's scheme~\cite{tHooft:1981ht}. The Abelian-like monopole currents satisfy  Abelian conservation rules kinematically. Exact Abelian (but kinematical) symmetries appear in non-Abelian QCD if VNABI occurs.  This is hence an extension of the Dirac idea~\cite{Dirac:1931} of monopoles in Abelian QED to non-Abelian QCD. 
  In the framework of $SU(2)$ QCD, some interesting numerical results were obtained as summarized later in concluding remarks. Abelian and monopole dominances as well as the Abelian dual Meissner effect are seen beautifully without any additional gauge-fixing~\cite{Suzuki:2007jp,Suzuki:2009xy}.  Also, the existence of the continuum limit of this new kind of Abelian-like monopoles is shown  with the help of the block-spin renormalization group with respect to the Abelian-like monopoles. A beautiful scaling behavior showing the existence of the continuum limit are observed with respect to the monopole density~\cite{Suzuki:2017lco} and the infrared effective monopole action~\cite{Suzuki:2017zdh}. This standpoint proposed by the author is called as random Abelian (RA) scheme from now on.

It is the purpose of this note is to discuss $SU(3)$ gauge invariance (
or gauge-choice independence) and to check whether non-Abelian color confinement can be explained or not under these three standpoints (Abelian projection, Pisa and
 RA schemes) of the Abelian dual Meissner pictures. In the following, I consider $SU(2)$ for simplicity mainly.

\section{
't~Hooft's
 Abelian projection scheme~\cite{tHooft:1981ht}}\label{Sec2}
Let me start by shortly reviewing the old 't~Hooft's idea of Abelian projection.  Although there is no need of performing gauge-fixing in non-perturbative QCD, 't~Hooft introduced a partial gauge-fixing of $SU(2)$ into $U(1)$. It is possible to show that a $U(1)$ Abelian monopole appears at the point where the gauge-fixing matrix becomes singular. 

Although there are infinite ways to do such a partial gauge-fixing, let me perform a 
 partial-fixing $V(x)$ by diagonalizing an adjoint $SU(2)$ operator $X(x)$ composed of $A_\mu(x)$.  $X(x)$ can be parametrized as
\begin{eqnarray}
X&=& \left(
  \begin{array}{cc}
   \cos\alpha +i\sin\alpha\cos\theta  & \sin\alpha\sin\theta e^{i\phi}   \\
   -\sin\alpha\sin\theta e^{-i\phi} & \cos\alpha-i\sin\alpha\cos\theta   \\
  \end{array}
\right),\end{eqnarray}
\begin{eqnarray}
VXV^{\dag}&=&X_d=\cos\alpha +i\sigma_3\sin\alpha,
 \label{Eq2}
\end{eqnarray}
where $V$ is an $SU(2)$ gauge-fixing matrix:
\begin{eqnarray}
V&=&\left(
  \begin{array}{cc}
    \cos(\frac{\theta}{2})   & \sin(\frac{\theta}{2})e^{-i\phi}   \\
   -\sin(\frac{\theta}{2})e^{i\phi}    & \cos(\frac{\theta}{2})   \\
  \end{array}
\right). \label{Eq3}
\end{eqnarray}
Here its diagonal elements are chosen to be real using $U(1)$ unfixed. 
Since $V(x)$ is also an $SU(2)$ matrix, let us consider its transformation property under arbitrary $W(x)\in SU(2)$. The transformed matrix $V^W$ is defined as diagonalizing $X^W=WXW^{\dag}$. Since the eigenvalues in (\ref{Eq2}) 
are the same, one gets 
\begin{eqnarray}
&&V^WX^WV^{W\dag}=V^WWXW^{\dag}V^{W\dag} \nn \\
               &=&V^WWV^{\dag}X_dVW^{\dag}V^{W\dag}=X_d. \label{Eq4}
\end{eqnarray}
There remains still $U(1)$ degree of freedom $d$ and hence from (\ref{Eq4})
one gets in general
\begin{eqnarray}
V^W=dVW^{\dag}. \label{Eq5}
\end{eqnarray}

Then one finds that the following quantity
\begin{eqnarray}
\Phi(x)&\equiv &V^{\dag}(x)\sigma_3V(x) \label{Eq6}
\end{eqnarray}
transforms as an adjoint operator: 
\begin{eqnarray}
\Phi'(x)=W(x)\Phi(x)W^{\dag}(x). \label{Eq7}
\end{eqnarray}

 An arbitrary adjoint operator $X(x)$ is considered above. 
But in any other complicated gauge-fixing schemes such as MA, 
it is possible to get the above adjoint
operator $\Phi(x)$ in terms of the MA gauge-fixing  matrices. Namely, different gauge-fixings are characterized by an adjoint operator $\Phi^{\alpha}(x)$ which has a space-time point where the gauge-fixing matrix becomes singular. Here the superscript $\alpha$ denotes  a different way of partial gauge-fixings. 

The 't~Hooft tensor is defined by
\begin{eqnarray}
F_{\mu\nu}^{\alpha}(x)&=&\Tr(\Phi^{\alpha}(x)G^{\alpha}_{\mu\nu}(x))\nn \\
&&-\frac{i}{g}\Tr(\Phi^{\alpha}(x)[D_\mu^{\alpha}\Phi^{\alpha}(x),D^{\alpha}_\nu\Phi^{\alpha}(x)]), \label{Eq8}
\end{eqnarray}
where $G^{\alpha}_{\mu\nu}$ is a non-Abelian field strength and $D^{\alpha}_\mu=\partial_{\mu}-igA^{\alpha}_{\mu}$ is a covariant derivative in one of Abelian projection schemes. This is reduced to 
\begin{eqnarray}
F^{\alpha}_{\mu\nu}&=&\Tr(\partial_\mu(\Phi^{\alpha} A^{\alpha}_{\nu})-\partial_\nu(\Phi^{\alpha} A^{\alpha}_{\mu})\nn \\
&& -\frac{i}{g}\Tr(\Phi^{\alpha}[\partial_\mu\Phi^{\alpha},\partial_\nu\Phi^{\alpha}]).\label{Eq9}
\end{eqnarray}
For simplicity in the following, the superscript $\alpha$ is not written explicitly  except for necessary cases.
If one expresses the adjoint operator in terms of $\theta(x)$ and $\phi(x)$
in Eq.(\ref{Eq3}) , one gets
\begin{eqnarray}
\Phi&=&\hat{\Phi}^a\sigma_a, \ \hat{\Phi}^a=(\sin\theta\sin\phi,\sin\theta\cos\phi,\cos\theta).\label{Eq11}
\end{eqnarray}
Then after the partial gauge-fixing, the third component of the gauge field becomes a photon-like field with respect to the remaining $U(1)$:
\begin{eqnarray}
A_\mu^{3'}&=&\Tr(\sigma_3A'_\mu)\nn \\
&=&\hat{\Phi}^aA_\mu^a - (1-\cos\theta)\partial_\mu\phi\nn \\
&=&a_\mu-\frac{1}{g}\frac{1}{1+\hat{\Phi}^3}\epsilon_{ab3}\hat{\Phi}^a
(\partial_\mu\hat{\Phi}^b), \label{Eq12}\\
F_{\mu\nu}&=& \partial_\mu A_\nu^{3'}-\partial_\nu A_\mu^{3'}\nn \\
&=&\partial_\mu a_\nu-\partial_\nu a_\mu+h_{\mu\nu},\label{Eq13} \\
h_{\mu\nu}&=& -\frac{1}{g}\epsilon_{abc}\hat{\Phi}^a(\partial_\mu\hat{\Phi}^b)(\partial_\nu\hat{\Phi}^c), \label{Eq14}
\end{eqnarray}
where $a_\mu=\hat{\Phi}^aA_\mu^a$. The second term in (\ref{Eq12}) has a line-like singularity and (\ref{Eq14}) coming from 
 the term 
 gives rise to an Abelian monopole current satisfying the Dirac quantization condition. It is to be stressed  that \textit{the original gauge field $A_\mu$ is assumed to be regular as usual, so that $a_\mu$ is also regular}. If one performs a gauge transformation $W(x)$ to so-called unitary gauge in such a way as $\Phi(x)$ is diagonalized,  the second term in (\ref{Eq12}) and (\ref{Eq13}) disappears. But then the projected Abelian gauge field after the gauge transformation becomes singular and contains a line-like singularity of the Dirac type. But note that such a singularity comes from the singularity of the original gauge-fixing matrix.
  
Although the Dirac quantization condition is satisfied, the 
't~Hooft's Abelian projection scheme is gauge dependent, since the Abelian monopoles come from the singularity of the gauge-fixing matrix $V(x)$ depending on the gauge chosen.
Moreover, all Abelian charged states with respect to the remaining $U(1)$ are confined due to the Abelian dual Meissner effect, but neutral states belonging to nontrivial representations of $SU(2)$ such as an adjoint one are not confined based on this Abelian projection scenario. It is very unsatisfactory, since color confinement of QCD says that color singlets alone can survive in the physical state.

\section{Random Abelian scheme of color confinement due to VNABI without any additional gauge-fixings}\label{Sec3}
The present author proposed a completely different approach of Abelian monopoles without performing any additional gauge-fixings, assuming VNABI exists in continuum QCD\cite{Suzuki:2014wya}. 
%
\subsection{Equivalence of VNABI and Abelian-like monopoles}
VNABI is defined as
\begin{eqnarray}
J_\nu(x)=\frac{1}{2}\epsilon_{\mu\nu\alpha\beta}D_\mu G_{\alpha\beta}(x)
=D_\mu G^*_{\mu\nu}(x). 
\end{eqnarray}
An interesting fact was found that VNABI $J_\nu(x)$ is nothing but violation of Abelian-like Bianchi identities written by Abelian-like field strengths. Using the Jacobi identity $\epsilon_{\mu\nu\rho\sigma}[D_{\nu},[D_{\rho},D_{\sigma}]]=0$. one gets 
\begin{eqnarray*}
[D_{\rho},D_{\sigma}]&=&[\partial_{\rho}-igA_{\rho},\partial_{\sigma}-igA_{\sigma}]\\
&=&-igG_{\rho\sigma}+[\partial_{\rho},\partial_{\sigma}],
\end{eqnarray*}
where the second commutator term of the partial derivative operators can not be discarded in general, since gauge fields may contain a (line-like) singularity. The relation $[D_{\nu},G_{\rho\sigma}]=D_{\nu}G_{\rho\sigma}$ and the Jacobi identity leads us to
\begin{eqnarray}
D_{\nu}G^{*}_{\mu\nu}&=&\frac{1}{2}\epsilon_{\mu\nu\rho\sigma}D_{\nu}G_{\rho\sigma} \nn\\
&=&-\frac{i}{2g}\epsilon_{\mu\nu\rho\sigma}[D_{\nu},[\partial_{\rho},\partial_{\sigma}]]\nn\\
&=&\frac{1}{2}\epsilon_{\mu\nu\rho\sigma}[\partial_{\rho},\partial_{\sigma}]A_{\nu}
=\partial_{\nu}f^{*}_{\mu\nu}, \label{eq-JK}
\end{eqnarray}
where $f_{\mu\nu}$ is defined as $f_{\mu\nu}=\partial_{\mu}A_{\nu}-\partial_{\nu}A_{\mu}=(\partial_{\mu}A^a_{\nu}-\partial_{\nu}A^a_{\mu})\lambda^a/2$. Namely Eq.(\ref{eq-JK}) shows that the violation of the non-Abelian Bianchi identity $J_\mu$, if exists at all,  is equivalent to that of the Abelian-like Bianchi identities.
Abelian-like monopole currents $k_{\mu}$ without any gauge-fixing are denoted by  the violation of the Abelian-like Bianchi identities:
\begin{eqnarray}
k_{\mu}=\partial_{\nu}f^*_{\mu\nu}
=\frac{1}{2}\epsilon_{\mu\nu\rho\sigma}\partial_{\nu}f_{\rho\sigma}. \label{ab-mon}
\end{eqnarray}
Eq.(\ref{eq-JK}) shows that 
\begin{eqnarray}
J_{\mu}=k_{\mu}. \label{J-K}
\end{eqnarray}

\subsection{Gauge invariance of the RA scheme of color confinement due to Abelian-like monopole $k_\mu=J_\mu$ condensation}
Let us assume VNABI exists in physical states in continuum QCD. Then Abelian-like monopoles $k_{\mu}(x)$ also appear necessarily in  physical states. From the antisymmetric property of the Abelian-like field strength, one gets Abelian-like conservation conditions~\cite{Arafune:1974uy}:
\begin{eqnarray}
\partial_\mu k_\mu=0. \label{A-cons}
\end{eqnarray}
Then there exist 3 in $SU(2)$ and 8 in $SU(3)$ conserved magnetic charges. There are $U(1)_m^3$ magnetic (kinematical) symmetries in $SU(2)$ and $U(1)_m^8$ in $SU(3)$.   
\begin{enumerate}
  \item 
To prove color electric gauge invariance of this RA 
scheme,
 I start from $SU(2)$. To say about the Abelian dual Meissner effect, it is necessary to define a local $U(1)$ symmetry having a photon-like field from $SU(2)$. An arbitrary $SU(2)$ gauge transformation is expressed as 
\begin{eqnarray}
V(x)=e^{i\vec{\alpha}(x)\cdot\vec{\sigma}},
\end{eqnarray}
where $\vec{\alpha}(x)=(\alpha_1(x),\alpha_2(x),\alpha_3(x))$. There are three independent $U(1)$ expressed in terms of $\alpha_i$  among $SU(2)$. 
Consider for example, only 
a gauge transformation
 expressed by a vector $\vec\alpha(x)=(0,0,\alpha_3(x))$, Then the photon-like $U(1)$ gauge field is played by $A_\mu^3(x)$ (Case1). $U(1)$ electric charges are 
characterized through the coupling of the Abelian gauge field.  For example, 
quarks in the fundamental representation have color electric charges $\pm g/2$ and 
gluons have $(-g, 0, +g)$ charges with respect to the chosen $U(1)$. One can express
quarks in the fundamental representation as
\be
q=\left(
\begin{array}{c}
u_3\\
d_3
\end{array}
\right),
\ee
where $u_3$ and $d_3$ have an electric charge $\pm g/2$ under the chosen $U(1)$.
With respect to magnetic monopoles, as shown above, there exist three Abelian-like conserved monopoles in $SU(2)$ with a magnetic charge:
\begin{eqnarray}
g_m^a=\int d^3x k_4^a(x).
\end{eqnarray}
Consider the case of $U_m(1)$ in $a=3$ direction as above, As Dirac suggested, if the quantization condition between any electric charge $Q$ and the magnetic charge $g_m$  is 
satisfied in such a way as
\begin{eqnarray}
Qg_m=2\pi n,
\end{eqnarray}
where $n$ is an integer, a quantum theory in Case1 could be formulated~\cite{Dirac:1931}.
However, 
whether the Dirac quantization condition is satisfied or not is not known a priori in our formulation on the contrary to the above
 't~Hooft's Abelian projection scheme. It must be proved rigorously by any other method for example, by adopting lattice QCD simulations.
If the Abelian monopoles satisfying the Dirac quantization condition exist in the continuum limit, all states having the above Abelian color charges are confined due to the Abelian Meissner effect caused by the condensation of
 $k_\mu^3(x)$.
 What happens about Abelian neutral states, say a neutral state $\bar{u}_3u_3-\bar{d}_3d_3$ in the adjoint representation? Non-Abelian color confinement says all states except for color singlets must be confined. In the $SU(2)$ meson case, $\bar{u}_3u_3+\bar{d}_3d_3$ can appear in physical states. Such a state as $\bar{u}_3u_3-\bar{d}_3d_3$ must be confined.

Here to be noted, one can consider any other $U(1)$ direction equally, since here no additional gauge-fixing as done by 't~Hooft~\cite{tHooft:1981ht} is introduced.
 Hence this idea is called as random Abelian scheme of color confinement.
For example, let us next pay attention to $U(1)$ in $a=1$ direction satisfying $\vec\alpha(x)=(\alpha_1(x),0,0)$ (Case 2). Then $A_\mu^1(x)$ becomes an Abelian gauge field and color electric charges are defined through the coupling of 
$A_\mu^1(x)$. The quark fields $(u_1,d_1)$ having an electric charge $\pm g/2$ in Case2 is expressed in terms of $(u_3,d_3)$ as 
\be
\left(
\begin{array}{c}
u_1\\
d_1
\end{array}
\right)=
\left(
\begin{array}{c}
\frac{(u_3+d_3)}{\sqrt{2}}\\
\frac{(u_3-d_3)}{\sqrt{2}}
\end{array}
\right).
\ee
Similarly, in $U(1)$ for $a=2$ direction satisfying $\vec\alpha(x)=(0,\alpha_2(x),0)$ (Case 3), The quark fields $(u_2,d_2)$ having an electric charge $\pm g/2$
with respect to $A_\mu^2(x)$ are shown as
\be
\left(
\begin{array}{c}
u_2\\
d_2
\end{array}
\right)=
\left(
\begin{array}{c}
\frac{(iu_3+d_3)}{\sqrt{2}}\\
\frac{(iu_3-d_3)}{\sqrt{2}}
\end{array}
\right).
\ee

The above neutral state in Case1 of the adjoint representation becomes a colored state with respect to $A_\mu^1(x)$ (Case2) and $A_\mu^2(x)$ (Case3). For example,
\begin{eqnarray}
\bar{u}_3u_3-\bar{d}_3d_3=\bar{u}_1d_1+\bar{d}_1u_1.
\end{eqnarray}
Hence if the monopole condensation occurs \textit{also in $a=1$ direction} with respect to $k_\mu^1(x)$, such a neutral state in Case1 must be confined. 
Only with respect to color singlets, one gets 
\begin{eqnarray}
\bar{u}_1u_1+\bar{d}_1d_1
=\bar{u}_2u_2+\bar{d}_2d_2
=\bar{u}_3u_3+\bar{d}_3d_3.
\end{eqnarray}

The above situations are more easily seen when one writes 
the quark-gluon coupling term in $SU(2)$ QCD as
\begin{eqnarray}
\bar{q}\gamma^{\mu}\frac{\sigma^a}{2}q A^a_{\mu}&=&\frac{1}{2} 
(\bar{u}_3\gamma_{\mu}d_3+\bar{d}_3\gamma_{\mu}u_3)A^1_{\mu} \nonumber\\
&&
-i\frac{1}{2}(\bar{u}_3\gamma_{\mu}d_3-\bar{d}_3\gamma_{\mu}u_3)A^2_{\mu}\nonumber\\
&&+\frac{1}{2}(\bar{u}_3\gamma_{\mu}u_3-\bar{d}_3\gamma_{\mu}d_3)A^3_{\mu}\label{u3d3}\\
&=&\frac{1}{2}(\bar{u}_1\gamma_{\mu}u_1-\bar{d}_1\gamma_{\mu}d_1)A^1_{\mu}\nonumber\\
&&
+\frac{1}{2}(\bar{u}_2\gamma_{\mu}u_2-\bar{d}_2\gamma_{\mu}d_2)A^2_{\mu}\nonumber\\
&&+\frac{1}{2}(\bar{u}_3\gamma_{\mu}u_3-\bar{d}_3\gamma_{\mu}d_3)A^3_{\mu},
\end{eqnarray}  
where the first equation (\ref{u3d3}) is expressed in terms of $u_3$ and $d_3$ alone.

When monopole condensation occurs in all three-color directions, physical states become $SU(2)$ color singlets alone. Hence non-Abelian $SU(2)$ color confinement is proved by means of  the dual Meissner effect caused by Abelian monopoles for all color channels. This mechanism was first found by the present author in  Ref~\cite{Suzuki:2009xy}. But at that time, the meaning of Abelian monopoles without doing any partial gauge-fixing was not clear.

\item
In $V(x)\in SU(3)$, the same situations hold. $V(x)$
is parametrized as
\begin{eqnarray}
V(x)&=&\exp{i\vec{\gamma}(x)\cdot\vec{\lambda}},
\end{eqnarray}
where $\vec{\gamma}(x)=(\gamma_1(x),...,\gamma_8(x))$ and $\lambda_i\ \ (i=1\sim 8)$
are the Gell-Mann matrices. One may consider any $U(1)$ with respect to 8 colors as a subgroup of $SU(3)$ corresponding to $U_m(1)^8$. But in $SU(3)$, there are two commutable $U(1)$ as the maximal torus group  and so one can consider a commutable $U(1)^2$ at the same time.

In the following, I consider $U(1)^2$ composed of $(\gamma_3(x),\gamma_8(x))$ 
alone as usual.
Color electric charges are characterized by the weight vectors. 
The quark field is regarded as a fundamental representation of 
SU(3) group. This has a form
\be
q=\left(
\begin{array}{c}
q_1\\
q_2\\
q_3
\end{array}
\right),
\ee
where the labels 1, 2, 3 correspond to the three types of the 
color-electric charge red ($R$), blue ($B$) and green ($G$).
By using the relation
\be
\vec{H}=(T_3,T_8)=
\left(
\begin{array}{ccc}
\vec{w}_1 & 0         & 0\\
0         & \vec{w}_2 & 0\\
0         & 0         & \vec{w}_3 
\end{array}
\right),
\ee
where $\vec{w}_{j}$ are the weight vectors of the SU(3) algebra,
\bea
\vec{w}_1&=& \left (\frac{1}{2}, \frac{1}{2\sqrt{3}} \right ),\quad 
\vec{w}_2= \left(-\frac{1}{2}, \frac{1}{2\sqrt{3}} \right ),\quad \\
\vec{w}_3&=& \left (0, -\frac{1}{\sqrt{3}} \right ),
\eea
one obtains an explicit form of the quark current,
\be
\vec{j}_{\mu}=g \sum_{j=1}^3 \vec{w}_j \; \bar{q}_j \gamma_{\mu} q_j.
\label{eqn:cur-general}
\ee
One finds that the color-electric charge is given by $g\vec{w}_j$~\cite{Koma:2001}.

The color-magnetic charge of the monopole is 
defined by $g_m \vec{\epsilon}_i$, where $\vec{\epsilon}_i$ are the root 
vectors of the SU(3) algebra,
\bea
\vec{\epsilon}_1&=&\left (-\frac{1}{2},\frac{\sqrt{3}}{2} \right ),\quad 
\vec{\epsilon}_2=\left(-\frac{1}{2},-\frac{\sqrt{3}}{2} \right ), \quad \\ 
\vec{\epsilon}_3&=&\left (1,0 \right ),
\eea
where the labels 1, 2, 3 correspond to 
dual red (${}^{*\!}R$), dual blue (${}^{*\!}B$) and 
dual green (${}^{*\!}G$).  Here, `` ${}^{*}$ '' denotes dual.
Both the gauge coupling $g$ and the dual gauge coupling $g_m$ 
are related by the Dirac quantization condition $gg_m=4\pi$.
It might be worthwhile to remember that the relation of 
the root vector and the weight vector of the SU(3) algebra is given by
\bea
\vec{\epsilon}_i \cdot \vec{w}_j &=&
\frac{1}{2} \left (\begin{array}{ccc}
 0 & 1 &  -1 \\
 -1 &  0 & 1 \\
1 &  -1 &  0
\end{array}\right )\\
&=&
\frac{1}{2}\sum_{k=1}^3 \epsilon_{ijk}
\equiv
\frac{1}{2} m_{i j},
\label{eqn:charge-relation}
\eea
where $m_{ij}$ is an integer which takes 0 or $\pm 1$.

As done in $SU(2)$ case, one can choose any $U(1)^2$ from $SU(3)$. For example, 
one can take $U(1)^2$ which is expressed by $(\lambda_4, (\sqrt{3}\lambda_3-\lambda_8(x))/2)$.

If one could prove that for all Abelian-like monopoles for $(a=1\sim 8)$ have the continuum limit satisfying the Dirac quantization condition, non-Abelian color confinement of $SU(3)$ is proved by means of the Abelian dual Meissner effect. Only $SU(3)$ color singlet states can appear in the physical space.
\end{enumerate}

\subsection{Proof in the framework of lattice field theory}
As shown above, the key point is whether such Abelian-like monopoles exist in the continuum limit, satisfying the Dirac quantization condition or not. As a rigorously reliable non-perturbative method of QCD, lattice studies seem to be unique. However, the topology of lattice QCD is different from that of continuum QCD.
To discuss such an operator which are non-zero only at singular points
is impossible within the framework of lattice field theories having finite degrees of freedom.

However, the equality (\ref{J-K}) shows that it is possible to study VNABI through Abelian-like monopoles on lattice following DeGrand-Toussaint~\cite{DeGrand:1980eq} as done in compact QED.
They define lattice monopoles through the Dirac quantization condition. Hence if one could prove that such monopoles are important and have the continuum limit clearly using various methods, say the Elitzur theorem~\cite{Elitzur:1975} or scaling studies using monopole block-spin transformation\cite{Shiba:1994db,Suzuki:2017lco,Suzuki:2017zdh}, 
VNABI $J_\mu$ exists and plays an important  role in color confinement of QCD. Also the Dirac quantization condition is proved to be satisfied at the same time.

On lattice, it is however impossible to distinguish between 't~Hooft's and VNABI types of monopoles from the beginning, since the only method known to extract lattice monopoles is at present that of Ref.\cite{DeGrand:1980eq}. It is the present 
author's opinion  that (1) when lattice monopoles without any additional gauge-fixing can explain confinement phenomena and have the continuum limit 
or (2) when monopoles after a partial gauge-fixing can explain confinement for any choice of gauges, namely, \textit{gauge-independently} in the continuum limit, one could say that they correspond to the Abelian-like monopoles of VNABI considered here. 

As shortly summarized in the last section\ref{Sec5}, the above authors' standpoint seems to work well at least in the framework of  $SU(2)$ QCD. 
  
\section{Pisa confinement scheme}\label{Sec4} 
In 2010, Bonati et al.~\cite{Bonati:2010tz,Bonati:2010B} 
found an interesting
 relation between violation of non-Abelian Bianchi identity (VNABI)  and the divergence of the 't~Hooft tensor $F^{\alpha}_{\mu\nu}$ corresponding to Eq.(\ref{Eq9}) in any 't~Hooft-type Abelian projection  represented in terms of $\Phi^{\alpha}(x)$, that is,
\begin{eqnarray}
\Tr(\Phi^{\alpha}(x) J^{\alpha}_\nu(x))&=&\Tr(\Phi^{\alpha}(x) D^{\alpha}_\mu G^{\alpha *}_{\mu\nu}(x))\nn \\
&=&\partial_\mu F^{\alpha *}_{\mu\nu}. \label{Eq16}
\end{eqnarray}

Based on this interesting relation, they
%
asserted that 
the gauge invariance of any 
't~Hooft's
 Abelian projection 
scheme is proved.
%

This 
assertion
 is derived from the following way:
\begin{enumerate}
\item When Lorentz invariance and Coleman-Mandula theorem\cite{Coleman:1967} are used, Ref\cite{Bonati:2010tz} says that it is possible to diagonalize $J_\mu(x)$ and $G_{\mu\nu}(x)$ transforming as an adjoint operator simultaneously using a unitary matrix $U(x)$
diagonalizimg $\Phi(x)$::
\begin{eqnarray}
(\Phi)_d(x)&=&U(x)\Phi(x)U^{\dag}(x), \\
(J_\nu)_{d}(x)&=&U(x)J_\nu(x)U^{\dag}(x), \label{JD}\\
(G^*_{\mu\nu})_{d}(x)&=&U(x)G^*_{\mu\nu}(x)U^{\dag}(x),\label{GD}
\end{eqnarray}
where the subscript $d$ stands for a diagonalized matrix.
\item Consider a matrix $\Phi_0(x)$ representing the fundamental weight. In $SU(2)$, it is $\sigma_3/2$. Then one gets 
\begin{eqnarray}
\Tr(\Phi_0(D_\mu G^*_{\mu\nu})_d)&=&\Tr(\Phi_0(J_\nu)_d).\label{PJ}
\end{eqnarray}
\item Consider the exact classical monopole solution\cite{tHooft:1974,Polyakov:1974} in the framework of 't~Hooft-Polyakov model containing a Higgs scalar field. After going to the unitary gauge, the direction of the Higgs field solution  becomes $\sigma_3/2$ which is equal to the above $\Phi_0(x)$. Then the gauge field solution becomes
\begin{eqnarray}
A_i^a(x)&=&\delta_{a3}\epsilon_{3ij}\frac{x_j}{r(r-x_3)},\label{TP}
\end{eqnarray}
which has a line-like singularity leading to the monopole. By direct calculations, the solution 
is found to satisfy
 the MA gauge condition\cite{Kronfeld:1987ri,Kronfeld:1987vd}:
\begin{eqnarray}
\partial_\mu A_\mu^{\pm}\pm ig[A^3_\mu,A^\pm_\mu]=0.  
\end{eqnarray}
Therefore Eq.(\ref{PJ}) corresponds to MA gauge in this classical solution case. 
\item The authors in \cite{Bonati:2010tz} go to the conclusion that beyond the classical solution the same situations exist dynamically in QCD without a Higgs field.
Hence $\Phi_0$ must correspond to the MA gauge\cite{Kronfeld:1987ri,Kronfeld:1987vd} in  the 
't~Hooft's Abelian projection scheme.
\item Since any Abelian projection scheme is represented by an adjoint operator $\Phi^{\alpha}(x)$ as shown in (\ref{Eq7}),  
they
 are related by a unitary transformation $U(x)$ as
\begin{eqnarray}
\Phi^{\alpha}(x)&=&U(x)^{\dag}\Phi_0(x)U(x),\label{Eq18}\\
J^{\alpha}_\nu(x)&=&U(x)^{\dag}(J_\nu)_d(x)U(x),\label{UJ}\\
G^*_{\mu\nu}(x)&=&U(x)^{\dag}(G^*_{\mu\nu})_{d}(x)U(x). \label{UG}
\end{eqnarray}
Hence the gauge invariance of 't~Hooft's Abelian projection scheme 
is proved.
\end{enumerate}

There are 
%
serious and unjustified points in the above scenario.  
The most serious is that, when 
showing gauge invariance of the 't~Hooft's scheme,
  they say that VNABI $J_\mu(x)$ and non-Abelian field strengths $G_{\mu\nu}(x)$ for all space-time directions can be simultaneously diagonalized in color space in terms of the same unitary matrix $U(x)$ when Lorentz invariance and Coleman-Mandula theorem\cite{Coleman:1967} are used. But it is not correct, since such a simultaneous diagonalization can be done \textit{only when} 
\begin{eqnarray}
[G_{\mu\nu}(x),G_{\alpha\beta}(x)]=0 \label{JJ}
\end{eqnarray}
for $(\mu\nu)\neq(\alpha\beta)$ in the case of $G_{\mu\nu}(x)$. Such a relation (\ref{JJ}) means that all $G_{\mu\nu}(x)$ for $(\mu,\nu)=1\sim 4$ are parallel in color space. 
These do not hold in general. 
This is seen easily when use is made of the classical hedgehog solution\cite{tHooft:1974, Polyakov:1974} as an example: 
\begin{eqnarray}
\hat{\Phi}^a&=&\frac{x_a}{r},\nn\\
A_0^a&=&0, \nn \\
A_i^a&=&\epsilon_{aij}\frac{x_j}{gr^2}.\nn
\end{eqnarray}
Non-Abelian field strengths are given by 
\begin{eqnarray}
gG_{ij}^a&=&\epsilon_{ajk}(\frac{\delta_{ki}}{r^2}-\frac{2x_kx_i}{r^4})
-\epsilon_{aik}(\frac{\delta_{kj}}{r^2}-\frac{2x_kx_j}{r^4})\nn \\
&&+\epsilon_{ijk}\frac{x_ax_k}{r^4}.
\end{eqnarray}
In the matrix form, one gets 
\begin{eqnarray}
[G_{ij},G_{kl}]=iG_{ij}^aG_{kl}^b\epsilon_{abc}\frac{\sigma_c}{2}.\label{GG}
\end{eqnarray}
Choose for example $(i=1,j=2,k=1,l=3)$. Then the following two are non-vanishing:
\begin{eqnarray}
gG_{12}^3&=&-\frac{x_3^2}{r^4},\nn \\
gG_{13}^2&=&\frac{x_2^2}{r^4},\nn
\end{eqnarray}
so that  RHS of Eq.(\ref{GG}) becomes $(i(x_2x_3)^2)/(2g^2r^8) \sigma_1\neq 0$.

  Beyond the classical solution level, it is possible to show that the simultaneous
diagonalization of $J_\mu^\alpha(x)$ leads to a contradiction. If the simultaneous
diagonalization hold good, then VNABI for two different partial gauge-fixings $\alpha\neq\beta$ are related by a unitary transformation $\hat{U}(x)$:
\begin{eqnarray}
J_\mu^\alpha(x)&=&\hat{U}(x)J_\mu^\beta(x)\hat{U}^{\dag}(x),\\
\Phi^\alpha(x)&=&\hat{U}(x)\Phi^\beta(x)\hat{U}^{\dag}(x).
\end{eqnarray}
Then both 't~Hooft tensors become equal from (\ref{Eq16}). This means that Abelian monopoles appearing after the Abelian projection for both cases are the same. This would lead to a contradiction.
Take an example of the so-called Polyakov gauge where Polyakov loops are chosen as the adjoint operator $\Phi(x)$.  It is shown \cite{Maxim:2003}
 rigorously 
 that in this gauge, Abelian monopoles are always time-like, i.e.,  static. Hence if Abelian monopoles in any Abelian projection scheme are the same, Abelian monopoles in any scheme are always static. Static Abelian monopoles can not contribute to usual Abelian Wilson loops corresponding to a pair of static quarks. The Abelian dual Meissner effect could not occur in any 
't~Hooft's Abelian projection scheme.  In addition, it is impossible to understand their assertion that the MA gauge has a preferred position among infinite Abelian projection schemes.

Hence in conclusion, 
the relation (\ref{Eq16}) 
 alone can not account for 
 the
  gauge invariance of the 
  't~Hooft's scheme.
 
\section{Concluding remarks}\label{Sec5}
In this note, gauge invariance of three typical standpoints on the Abelian dual Meissner effect presented so far are discussed. The Pisa scheme is not justified and  only the RA idea\cite{Suzuki:2014wya,Suzuki:2017lco} based on VNABI without performing any 
partial  gauge-fixing is shown to prove gauge invariance, if rigorous lattice studies could show the existence of such Abelian-like monopoles satisfying the Dirac quantization condition in the continuum limit.

Finally, to show the picture based on Abelian-like monopoles is very promising, 
let me summarize shortly numerical $SU(2)$ data\cite{Suzuki:2007jp,Suzuki:2009xy,Suzuki:2017lco,Suzuki:2017zdh}  
taken by the present author and his collaborators. For the details, refer to these original papers. 
\subsection{Perfect Abelian and perfect monopole dominances and the dual Meissner effect} 
As discussed above, one of possible ways to prove the importance of the Abelian-like monopoles from VNABI is to perform numerical lattice studies without introducing any partial gauge-fixing. Such studies were done in Refs\cite{Suzuki:2007jp,Suzuki:2009xy}. But at that time  the theoretical background of the Abelian monopoles used without any partial gauge-fixing were not known. Although no partial gauge-fixing is done, one can extract Abelian link field as done in MA gauge. 
\begin{enumerate}
  \item As expected from the Abelian dual Meissner effect, perfect Abelian dominance, namely, $\sigma_{full}=\sigma_a$ is proved with the help of the powerful noise reduction multilevel method\cite{Luscher:2001up}. 
  \item Perfect monopole dominance, namely, $\sigma_{full}=\sigma_a=\sigma_m$ is shown by using so many vacuum ensembles with the help of additional random gauge transformations. Asymptotic scaling and volume independence were studied also. 
\item The dual Meissner effect due to the Abelian monopole condensation is also observed very beautifully. 
All cylindrical components of the color-electric fields 
${\cal O}(s)=E_{{\rm A}i}(s)=\bar{\Theta}_{4i}(s)$ are measured.
It is found that only $E_{{\rm A}z}$ has correlation with the Wilson loop.
and that the penetration depth~$\lambda$ shows the good scaling 
behavior.
\item (The dual Amp\`ere law)
To see what squeezes the color-electric field, 
the Abelian (dual) Amp\`ere law was studied: 
\begin{eqnarray}
\vec{\nabla}\times\vec{E}=
\partial_{4}\vec{B}+2\pi\vec{k}\;, 
\end{eqnarray}
where $\vec{E}$ and $\vec{B}$ are Abelian electric and magnetic fields, respectively.
The correlation of each term with the Wilson loop 
is evaluated on the mid-plane of the $q$-$\bar{q}$ system. 
It was found that only the azimuthal components are non-vanishing.  If the color-electric field is purely of the Coulomb type, 
the curl of the electric field is zero. 
On the contrary, the curl of the electric field
is non-vanishing and is reproduced by the monopole currents.
In any case, the dual Amp\`ere law is satisfied, which 
is a clear signal of the Abelian dual Meissner effect.
This result is quite the same as those
observed in the MA gauge~\cite{Koma:2003gq,Koma:2003hv}. 
\item (The coherence length)
The coherence length was evaluated from  
the correlation function between
the squared monopole density~${\cal O}(s)=k_{\mu}^2(s)$
and the Wilson loop~\cite{Chernodub:2005gz}. It was found that 
the coherence lengths determined show the scaling behavior as a function of lattice 
spacing~$a(\beta)$.
\item (The vacuum type)
Taking the ratio of the penetration depth 
and the coherence length, the Ginzburg-Landau (GL) parameter 
$\sqrt{2}\kappa=\lambda/\xi$ can be estimated,
which characterizes the type of the superconducting vacuum.
It was found  that the GL parameters determined show the scaling behavior
and the values are all about one.
This means that the vacuum type of $SU(2)$ QCD is near the border 
between the type~1 and~2 dual superconductor.
\end{enumerate}

\subsection{Block-spin transformation studies of Abelian monopoles}
To prove such Abelian-like monopoles of VNABI exist actually in the continuum limit, it is important to study the continuum limit of the density and the effective action of such Abelian-like monopoles more rigorously.  These studies were done in Refs\cite{Suzuki:2017lco,Suzuki:2017zdh}.
But such a quantity like the monopole density is always positive definite. 
On lattice, it is well known that many lattice artifact monopoles having no continuum limit appear in the thermalized vacuum. Such lattice artifact monopoles contribute
to the measurement of the monopole density and the monopole effective action. 
It is absolutely necessary to make the vacuum as smooth as possible so that lattice artifact effects may be reduced substantially. Four different partial gauge-fixings smoothing the vacuum were introduced to check gauge choice independence. They are MA\cite{Kronfeld:1987ri,Kronfeld:1987vd}, Maximal center gauge (MCG)~\cite{DelDebbio:1996mh,DelDebbio:1998uu}, Laplacian center gauge (LCG)~\cite{Faber:2001zs} and maximal Abelian Wilson loop gauge (MAW) where $1\times 1$ Abelian Wilson loop is maximized.  A block-spin transformation of lattice monopoles\cite{Ivanenko:1991wt, Shiba:1994db} is introduced also. The tadpole improved action was adopted for $\beta=3.0\sim 3.9$.   
The primary monopole $k_\mu(s)$ is defined on a $a^3$ cube and the $n$-blocked monopole
$k_\mu^{(n)}(s_n)$ is defined on a cube 
with a lattice spacing $b=na$. $s_n$ is the site on the reduced blocked lattice.
\begin{eqnarray}
k_{\mu}^{(n)}(s_n) = \sum_{i,j,l=0}^{n-1}k_{\mu}(ns_n+(n-1)\hat{\mu}+i\hat{\nu}
     +j\hat{\rho}+l\hat{\sigma}). \nn
\end{eqnarray}

Using the blocked monopole currents, the density was measured in Ref.\cite{Suzuki:2017lco}.
The density is defined as follows:
\begin{eqnarray}
\rho(\beta,n)=\frac{\sum_{\mu,s_n}\sqrt{\sum_a(k_{\mu}^a(s_n))^2}}{4\sqrt{3}V_nb^3},\label{RHO}
\end{eqnarray}
where $V_n=V/n^4$ is the 4-dimensional volume of the reduced lattice, $b=na(\beta)$ is the spacing of the reduced lattice after $n$-step blockspin transformation. Here the superscript $a$ denotes a color component. Note that $\sum_a(k_{\mu}^a)^2$
is gauge-invariant in the continuum limit. In general, the density $\rho$ (\ref{RHO}) is a function of two variables $\beta$ and $n$. But if $\rho$ is plotted versus $b=na(\beta)$, a beautiful universal scaling function was obtained:
\begin{eqnarray}
\rho(\beta,n)\to \rho(b=na(\beta)).
\end{eqnarray}  
 Clear scaling behaviors $\rho(b)$ up to the 12-step blocking transformations for $\beta=3.0\sim 3.9$ are seen. Hence for fixed $b$, if one takes $n\to \infty$, $a(\beta)\to 0$, that is, the continuum limit is obtained. In addition to the scaling behaviors, the obtained scaling function is the same for four different gauges adopted. Gauge independence is naturally expected in the continuum limit.

   Similar results showing beautiful scaling behaviors were obtained with respect to the renormalization flow of the monopole effective action\cite{Suzuki:2017zdh}:
\begin{eqnarray}
S_{eff}(\beta,n) \to S_{eff}(b=na(\beta)).
\end{eqnarray}
    
\subsection{$SU(3)$ studies}   
   Numerical calculations in $SU(3)$, although still imperfect, were done recently without any additional partial gauge-fixings\cite{Suzuki:2021A,Suzuki:2021B, Ishiguro:2022}. Perfect Abelian dominance is obtained also with the use of the multilevel method\cite{Luscher:2001up} for three different $\beta$ but still lattice volumes are small. Perfect monopole dominance and the Abelian dual Meissner effect are also observed as in $SU(2)$, but only a case for $\beta=5.6$ of the WIlson action on $24^3\times 4$  was studied owing to the fact that tremendous number of vacuum ensembles in $SU(3)$ were needed. Further intensive studies are necessary to prove the picture based on the new type of monopoles from VNABI in $SU(3)$. 

\vspace{.3cm}
   
\section{Acknowledgements}
This work used High Performance Computing resources provided by Cybermedia Center of Osaka University through the HPCI System Research Project (Project ID: hp210021). The numerical simulations of this work were done also using High Performance Computing resources at Research Center for Nuclear Physics  of Osaka University, at Cybermedia Center of Osaka University, at Cyberscience Center of Tohoku University and at KEK. The author would like to thank these centers for their support of computer facilities. This work was financially supported by JSPS KAKENHI Grant Number JP19K03848.
 The author is very grateful to T.Kugo for the discussions concerning simultaneous diagonalization of all four components of VNABI and K.Ishiguro for the long-term collaboration  on  numerical works.

\end{document}